\pdfoutput=1
\documentclass[10pt]{iopart}

\expandafter\let\csname equation*\endcsname\relax
\expandafter\let\csname endequation*\endcsname\relax
\usepackage{amsmath}
\usepackage{amssymb}

\usepackage{graphicx}
\usepackage{xcolor}
\usepackage{bm}
\usepackage{dcolumn}
\usepackage{comment}
\usepackage{placeins}
\usepackage{cite}

\usepackage[hidelinks]{hyperref}

\begin{document}

\title[QS EFC for ITER]{Quasi-symmetric error field correction and applications to ITER}

\author{G.-G. Seo$^{1}$, J.-K. Park$^{1,*}$, S. J. Han$^{1}$ and M. Dubrov$^{2}$}

\address{$^{1}$Department of Nuclear Engineering, Seoul National University,
Seoul 08826, Republic of Korea}
\address{$^{2}$ITER Organization, Route de Vinon-sur-Verdon, CS 90 046, 13067 St Paul Lez Durance Cedex, France}
\address{$^{*}$Author to whom any correspondence should be addressed.}

\ead{jkpark@snu.ac.kr}

\begin{abstract}
Reliable correction of nonaxisymmetric error fields (EFs) is essential to the safety and performance of tokamak operation. Although resonant error field correction (EFC) is well established, supported by an improved understanding of 3D plasma response, the residual fields left after resonant EFC remain an open question: they are expected to be predominantly nonresonant, yet can still degrade both confinement and stability. Here we introduce a systematic EFC scheme that minimizes resonant and nonresonant EF effects simultaneously, based on neoclassical torque response computed from self-consistent perturbed equilibria. The scheme extends the method developed for designing quasi-symmetric magnetic perturbations to the case where actual error fields are present. Application to standard ITER target plasmas with a range of intrinsic EF scenarios demonstrates the advantages of this quasi-symmetric (QS) EFC scheme for controlling residual EFs. QS EFC consistently yields low-torque solutions while strongly suppressing resonant response, outperforming single-mode resonant overlap EFC in most cases and approaching the performance of multimodal resonant EFC. We also show that the QS EFC solution varies only tolerably between half- and full-$I_p$ ITER scenarios, despite the greater sensitivity expected from its higher-order nature.
\end{abstract}

\vspace{2pc}

\submitto{Nuclear Fusion}
\maketitle
\ioptwocol

\section{Introduction}
\label{sec:intro}
Tokamak operation is highly sensitive to a small three-dimensional (3D) magnetic field that breaks toroidal symmetry. Even a perturbation at the level of $\delta B/B = 10^{-3}\sim 10^{-4}$ can trigger a dangerous resonant response such as locked modes or tearing modes~\cite{LaHayeEtAl1992,Boozer1996,Fitzpatrick1998,ParkJKPRL07,Finn2015,Fitzpatrick2023}. It can also fundamentally alter particle motion and reduce toroidal rotation~\cite{CallenNF09,ParkJKPOP09_APS,CallenNF11}, thereby degrading overall confinement and stability. These small 3D fields are inevitable in tokamaks, due to imperfections in the construction and assembly of the primary magnets and surrounding machine structures. In particular, misalignments in the toroidal field (TF), poloidal field (PF), and central solenoid (CS) coils such as millimeter-level shifts and tilts, or coil windings and leads, can intrinsically produce such a level of 3D error fields (EFs)~\cite{GribovIAEA12,Bai2025}, risking tokamak performance and operation.

Error field correction (EFC) mitigates these risks by using additional nonaxisymmetric coils and fields that can effectively compensate intrinsic EFs~\cite{Park2008,Reimerdes2011,Bandyopadhyay2025}.
Because EFC is actuator-limited, the choice of how EF ``severity'' is quantified becomes decisive. The adopted metric effectively dictates which spectral structures the limited actuators will prioritize, e.g. the dominant overlap structure~\cite{Park2008,Pharr2024}. This quantification has been progressively improved and applied to the International Thermonuclear Experimental Reactor (ITER) EFC strategy. Earlier practice employed the ``3-mode'' criterion~\cite{Buttery1999,Scoville_2003} based on vacuum superposition in the estimation of resonant fields. This gave useful guidance but did not always isolate the plasma response structures that matter most for resonant drive and locking.

Recent approaches instead emphasize resonant physics via
plasma response modeling and coupling matrices that connect
external 3D fields to shielded resonant flux on selected rational
surfaces~\cite{Park2008}. In this framework, Singular Value Decomposition (SVD) schemes systematically identify dominant response structures and define device-agnostic overlap measures for ranking locking-relevant
content~\cite{Park2008,Park2012,Pharr2024}. With this scheme as a reference, the focus of EFC has moved from dominant-structure targeting (overlap) to multi-surface resonance minimization~\cite{LanctotPOP17}, and then to formulations that include nonresonant physics such as neoclassical toroidal viscosity (NTV) torque~\cite{Buttery2012,Park2021,YangSMNC24}.

A central issue is that resonant minimization alone does not uniquely control the residual 3D field spectrum.
Different coil solutions can achieve comparable resonant suppression while leaving substantially different nonresonant fields, which can still drive NTV braking and reduce toroidal rotation~\cite{ShaingNF10_general,ParkJKPRL09,Cole2008,Cole2011,Satake2011,ParkJKPRL13,LoganNF16,Park2017,KimKNF17}. The reduced toroidal rotation and its shear can in turn induce adverse effects on macro-scale instabilities such as tearing modes and resistive wall modes, as well as on micro-scale turbulence. Furthermore, the resonant field threshold itself against locked modes can be substantially reduced as a result of residual nonresonant error fields~\cite{ColePRL07,ReimerdesNF09}. A resonance-only optimum can therefore be operationally suboptimal even when resonant metrics appear satisfactory~\cite{Cole2008,GerhardtPPCF10,Buttery2012,Paz-SoldanNF15}.

To fill this gap, we propose a new EFC strategy for ITER based on the concept of quasi-symmetric magnetic perturbation (QSMP), which comprehensively minimizes NTV as validated in KSTAR and DIII-D~\cite{Park2021,YangSMNC24}. The major extension from the earlier QSMP validations is the inclusion of actual error field sources rather than their proxy, and its systematic application to various ITER scenarios. Note that ITER imposes a stringent operational constraint. The conventional compass scans~\cite{MenardNF10,Park2012,Paz-SoldanPOP14,InNF15} based on disruptive locked-mode response are not acceptable in the baseline scenario due to its high stored energy with the plasma current $I_p=15$~MA~\cite{GribovIAEA12,Pharr2024}.
We therefore adopt a low-risk workflow, in which optimization is performed in a reduced-current scenario and the resulting correction currents are applied at 15~MA without re-optimization~\cite{Pharr2024,PazSoldan2022}.

This work makes three contributions. First, we introduce a torque-coupling matrix formalism that maps each EF source to the actuator choice and polarity suited to it, so that actuators can be planned beyond resonance-only metrics~\cite{Park2021}. Second, we compare the operational overlap reference, a conditional multimodal resonance-optimal extension, and quasi-symmetric (QS) EFC, which uses freedom left by the resonant targets to minimize torque. The comparison covers ITER-relevant PF/TF/CS shifts and tilts. Third, we test the transfer of currents optimized at reduced current to the 15~MA baseline without re-optimization, complementing the methodology developed for non-disruptive EF identification~\cite{PazSoldan2022,hu2024non,Piron2024}.

\section{Torque-coupling matrix for extended EFC}
\label{sec:torque-matrix}

Conventional EFC has been developed primarily to suppress resonant magnetic perturbations, motivated by their immediate impact on disruptive response including locked modes~\cite{Park2008,Pharr2024}.
This resonance-centric perspective is also operationally transparent because of its bifurcating nature, in which the plasma is driven onto a disruptive path once the resonant drive exceeds a critical level. Two issues arise, however, when EFC is extended to include nonresonant perturbations. First, NTV is nonlinear in the perturbation, so it is difficult to assign weighting across different $\delta\vec{B}_{mn}$ components. Second, NTV effects are gradual rather than bifurcating, and act on performance indirectly through rotation rather than directly, so it is not straightforward how to quantify the optimal EFC criterion.

\begin{figure*}[t]
  \centering
  \includegraphics[width=\textwidth]{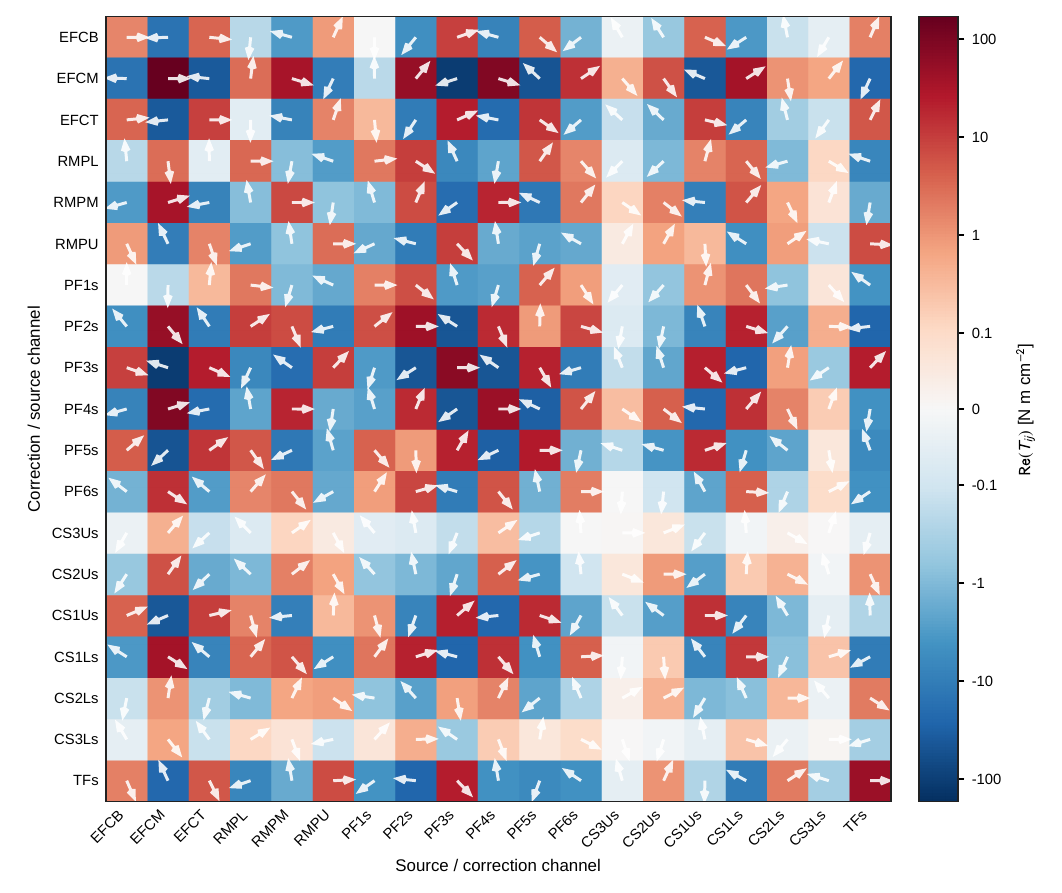}
  \caption{Full torque-coupling matrix $\mathbf{T}$ for the 15~MA ITER baseline scenario, evaluated for unit shifts of the TF/PF/CS coils as EF sources and for the six correcting-coil families EFCB/EFCM/EFCT/RMPL/RMPM/RMPU. Rows and columns both run over correctors and sources, so the off-diagonal blocks are the source--corrector coupling $\mathbf{T}_{cs}$ and its Hermitian conjugate $\mathbf{T}_{sc}$, while the diagonal blocks $\mathbf{T}_{cc}$ and $\mathbf{T}_{ss}$ give the corrector and source self-terms. Color shows $\Re(T_{ij})$ in N\,m\,cm$^{-2}$ using a symmetric logarithmic scale. White arrows show the complex phase of every matrix element. For a positive-real source amplitude, the torque-reducing corrector phase in the source--corrector block is opposite to the displayed coupling-phase arrow.}
  \label{fig:torque-matrix}
\end{figure*}

One approach is to minimize the total NTV torque integrated over the plasma, using the set of available control coils. One can estimate the integrated NTV torque self-consistently with perturbed equilibria by following the method of the Generalized Perturbed Equilibrium Code (GPEC)~\cite{ParkJKPOP07,LoganPOP13,Park2017}, including the non-adiabatic anisotropic pressure that is the origin of NTV~\cite{ParkJKPRL09,ParkJKPOP11}. According to this method, there is a unique answer to the optimal EFC. The integrated torque $T$ is expressed as
\begin{equation}
T(\mathbf{I}) = \mathbf{I}^\dagger \mathbf{T} \mathbf{I},
\label{eq:torque-quadratic}
\end{equation}
where $\mathbf{T}$ is a Hermitian matrix~\cite{Park2017,Park2021} and $\mathbf{I}$ is an augmented complex vector that represents correcting-coil currents and error field source amplitudes including their toroidal phases. $\mathbf{T}$ represents the contribution of each correction and source, as well as their nonlinear coupling, to the integrated torque, and is designated as the ``torque-coupling'' matrix throughout this paper. That is, the torque-coupling matrix contains all the information needed for the purpose of NTV minimization in our framework. This is highlighted in
Fig.~\ref{fig:torque-matrix}, which is constructed by GPEC for the ITER baseline scenario target. Once the torque-coupling matrix is fixed, the torque-minimizing currents follow directly. Without an EF source, the optimum is the eigenvector of the corrector block with the minimum eigenvalue, which defines the QSMP~\cite{Park2021}. With a given EF source, it is the minimum of the quadratic form with the source amplitudes held fixed. The details are as follows.

The augmented vector and matrix decompose as
\begin{equation}
\mathbf{I} \equiv
\begin{bmatrix}
\mathbf{I}_c \\
\mathbf{I}_s
\end{bmatrix},
\qquad
\mathbf{T} \equiv
\begin{bmatrix}
\mathbf{T}_{cc} & \mathbf{T}_{cs} \\
\mathbf{T}_{sc} & \mathbf{T}_{ss}
\end{bmatrix},
\label{eq:torque-block}
\end{equation}
where $\mathbf{I}_c$ denotes the complex correcting-coil current vector for the available actuator channels. In ITER, this includes three rows (bottom, middle, top) of ex-vessel EFC coils, EFCB, EFCM, and EFCT, and three rows (lower, middle, upper) of resonant magnetic perturbation (RMP) coils, RMPL, RMPM, and RMPU. $\mathbf{I}_s$ collects the amplitudes of the potential error field sources, for example unit shifts or tilts depending on the adopted normalization. The total torque is then
\begin{equation}
T(\mathbf{I}_c,\mathbf{I}_s)
=
\mathbf{I}_c^\dagger\mathbf{T}_{cc}\mathbf{I}_c
+ 2\,\Re\!\left[\mathbf{I}_c^\dagger\mathbf{T}_{cs}\mathbf{I}_s\right]
+ \mathbf{I}_s^\dagger\mathbf{T}_{ss}\mathbf{I}_s,
\label{eq:torque-expanded}
\end{equation}
where the cross-term represents the interference between the intrinsic error field and the applied correction field.
This term governs correction planning, since it determines, to leading order, whether a small applied current in a given actuator channel increases or reduces the torque associated with a specific error field source.
For fixed $\mathbf{I}_s$, Eq.~(\ref{eq:torque-expanded}) is quadratic in $\mathbf{I}_c$ and its minimum is
\begin{equation}
\mathbf{I}_c^\ast = -\,\mathbf{T}_{cc}^{-1}\,\mathbf{T}_{cs}\,\mathbf{I}_s ,
\label{eq:Ic-opt}
\end{equation}
provided that this solution lies within the coil current limits. Otherwise the minimum is found numerically on the boundary of the allowed currents.

In Fig.~\ref{fig:torque-matrix}, the real part of each matrix element is shown in N\,m\,cm$^{-2}$ with a symmetric logarithmic color scale.
Positive and negative real components are indicated in red and blue, respectively, while the white arrows show the complex phase of every matrix element.
For a positive-real source amplitude, the correcting-current phase that reduces the torque is shifted by $\pi$ from the displayed source--corrector coupling phase.
The figure can thus be read as an actuator--polarity map. For each EF source, it shows which coil family reduces the torque most effectively and which current phasing should be used.

\begin{figure*}[t]
  \centering
  \includegraphics[width=\textwidth]{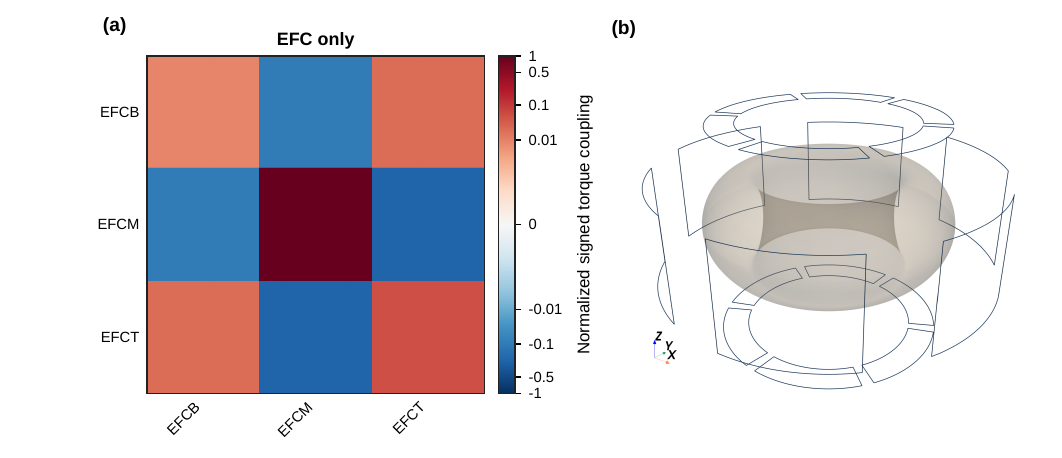}
  \caption{Torque-coupling matrix evaluated in the EFC-only actuator space for the 15~MA baseline equilibrium. Panel (a) uses a signed normalization to the largest absolute real element of the selected actuator matrix. Panel (b) shows the EFCB, EFCM, and EFCT actuator geometry.}
  \label{fig:qsmp-matrix-efc}
\end{figure*}

\begin{figure*}[t]
  \centering
  \includegraphics[width=\textwidth]{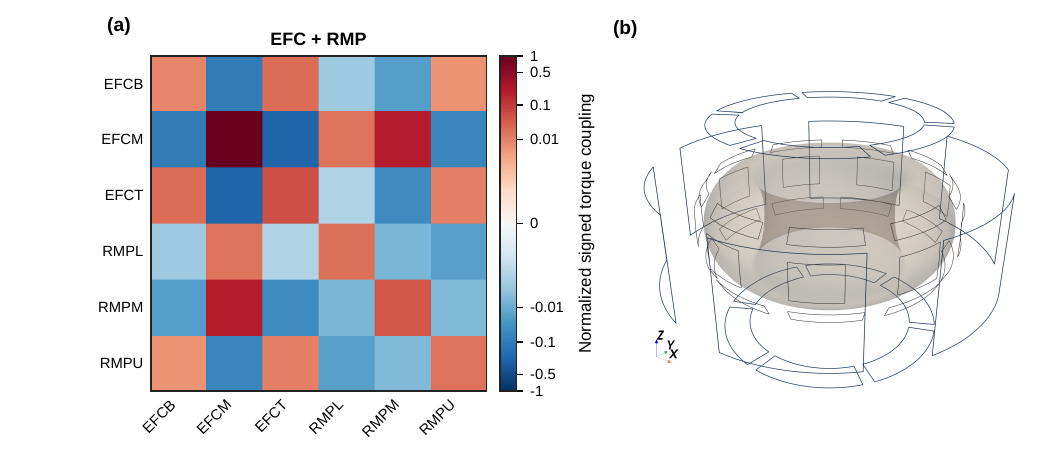}
  \caption{Torque-coupling matrix evaluated in the combined EFC and RMP actuator space for the 15~MA baseline equilibrium. The actuator set comprises EFCB/EFCM/EFCT and RMPL/RMPM/RMPU.}
  \label{fig:qsmp-matrix-efc+rmp}
\end{figure*}

\begin{figure*}[t]
  \centering
  \includegraphics[width=\textwidth]{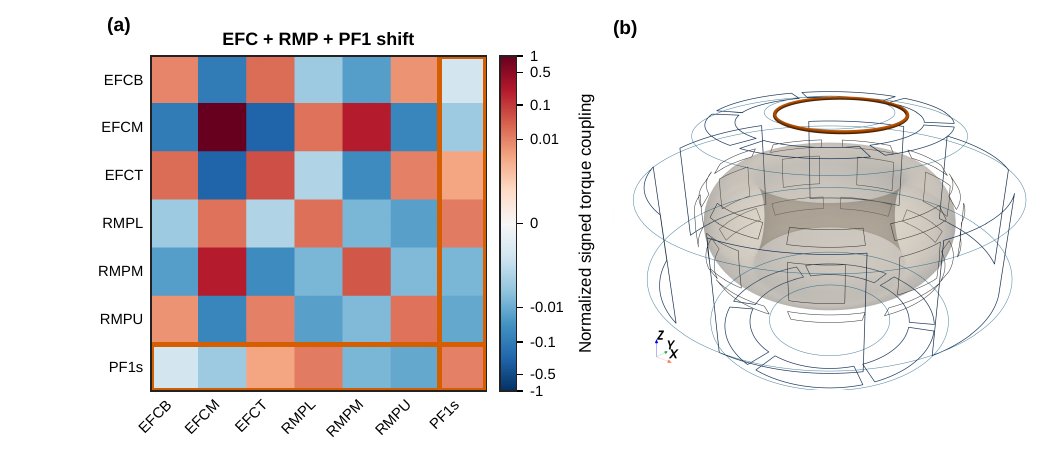}
  \caption{Torque-coupling matrix in the combined EFC and RMP actuator space augmented by a unit PF1-shift source for the 15~MA baseline equilibrium. The PF1 row and column are outlined in red.}
  \label{fig:qsmp-matrix-efc+rmp+pf1}
\end{figure*}

Figure~\ref{fig:qsmp-matrix-efc} shows the torque-coupling matrix when only the EFC coils are available as actuators.
In this three-channel space, the minimum-eigenvalue direction of $\mathbf{T}_{cc}$ prescribes how to combine the EFC currents to minimize the NTV torque. This is the idea of generating a quasi-symmetric magnetic perturbation with EFC coils alone~\cite{Park2021}.

When the actuator space is expanded to include RMP coils, as shown in Fig.~\ref{fig:qsmp-matrix-efc+rmp}, the minimum-eigenvector solution gains additional degrees of freedom.
The added actuators can give a larger torque reduction under the same current limits, or a comparable reduction with a smaller total current.
The matrix therefore allows different actuator sets to be compared and shows which channels contribute most to torque mitigation.

Figure~\ref{fig:qsmp-matrix-efc+rmp+pf1} is the case closest to actual EFC because it includes an intrinsic EF source in the augmented representation.
In this example, the error field is generated by a unit PF1 shift, represented by fixing the corresponding source amplitude to unity, $(\mathbf{I}_s)_{\mathrm{PF1}}=1$.
The augmented vector $\mathbf{I}=[\mathbf{I}_c;\mathbf{I}_s]$ then contains both the correcting-coil currents and the fixed error field amplitude, and the torque-coupling matrix captures their combined contribution to the NTV torque.
With the source amplitude fixed, Eq.~(\ref{eq:Ic-opt}) gives the correcting-coil current amplitudes and phases that minimize the torque in the presence of that error field. The PF1 column of the matrix supplies $\mathbf{T}_{cs}$ and the corrector block supplies $\mathbf{T}_{cc}$.
The strongest couplings in the matrix already indicate which actuator channels are likely to reduce the torque for the given source, before any full numerical optimization is carried out.

The same representation extends to cases with multiple EF sources.
Because the error field realization $\mathbf{I}_s$ is fixed for a given misalignment scenario, multiple sources can be included simultaneously without ambiguity, and the interpretation stays consistent within the same current-space basis.
This property matters in realistic configurations, where multiple PF, TF, and CS shifts or tilts are present simultaneously.
In such cases, the torque-coupling matrix still maps each EF source to an actuator choice and current phasing, and it also indicates the cost of correction.
Coils with strong coupling can reduce the torque considerably with a small current, whereas weakly coupled coils need a much larger current for the same effect. This is why specific EF sources repeatedly favor particular coil families in the QS EFC solutions.
\section{Applications to ITER EFC}
\label{sec:methods}

Section~\ref{sec:torque-matrix} introduced the torque-coupling matrix framework and its interpretation.
In this section, we describe how the framework is applied to ITER-relevant equilibria and intrinsic EF source sets, and how it is used to compare correction strategies under operational constraints.
We first introduce the equilibria and rational surfaces that define the resonant targets, then describe the control-surface representation used to provide a common boundary input to GPEC.
We next define the EF source set constructed from realistic PF/TF/CS misalignments, and we summarize the plasma-response metrics used to quantify resonance and torque.
Finally, we describe the three correction strategies considered in this work and present a low-risk protocol in which correction currents identified at reduced plasma current are transferred to the 15~MA baseline without re-optimization.

\begin{figure*}[t]
  \centering
  \includegraphics[width=\textwidth]{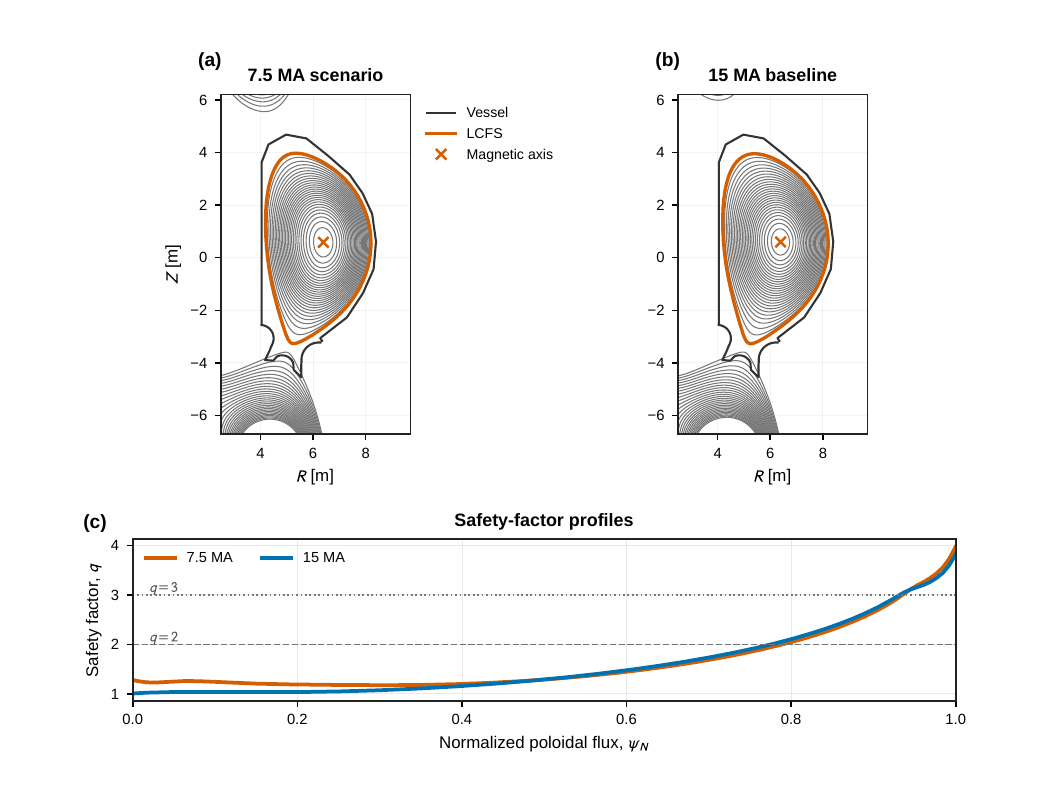}
  \caption{Equilibria and safety-factor profiles used in this work. Panels (a) and (b) show poloidal-flux contours, the last closed flux surface, vessel outline, and magnetic axis for the 7.5~MA and 15~MA cases, respectively. Panel (c) compares their $q$ profiles. Both have $q_{95}<4$, with the dominant $n=1$ rational surfaces including $q=2$ and $q=3$.}
  \label{fig:iter-bsl-eqinfo}
\end{figure*}

All calculations are performed for the ITER baseline-scenario equilibrium at $I_p=15$~MA and $B_T=5.3$~T and for a reduced-current case at $I_p=7.5$~MA and $B_T=2.65$~T.
The reduced-current equilibrium is used for EF identification and optimization at reduced disruption risk.
Figure~\ref{fig:iter-bsl-eqinfo} compares the $q$ profiles of the two equilibria.
For both cases, $q_{95}<4$ and the dominant $n=1$ rational surfaces include $q=2$ and $q=3$.
These surfaces are used consistently throughout the paper to define the resonant-response metrics and the correction targets.

To ensure a consistent boundary representation of applied perturbations, we define a fixed control surface located just outside the last closed flux surface, for example at a slightly larger normalized poloidal flux $\psi_N \gtrsim 1$.
All external perturbations are represented on the same control surface, including both intrinsic EF sources and correcting-coil fields.
On this surface, we evaluate the complex external spectrum denoted $\tilde{\Phi}_x$, which is used as the boundary input to GPEC.
GPEC then computes the shielded plasma response and returns the resonant (shielded) flux on selected rational surfaces, denoted $\tilde{\Phi}_r$.
This common mapping from $\tilde{\Phi}_x$ to $\tilde{\Phi}_r$ under a fixed equilibrium and boundary definition allows EF sources and correction strategies to be compared on the same footing.

\begin{figure}[t]
  \centering
  \includegraphics[width=\linewidth]{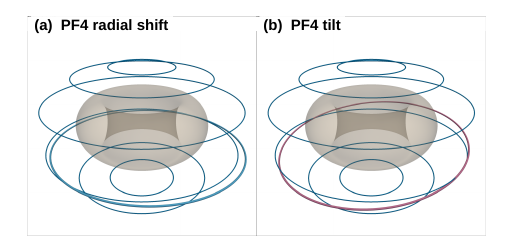}
  \caption{Schematic examples of (a) a PF4 radial shift and (b) a PF4 tilt. Transparent blue indicates the nominal PF-coil positions and the colored coil indicates the displaced geometry. The deformation is exaggerated for visibility and is not to scale.}
  \label{fig:pf4tilt-pf3shift}
\end{figure}

The intrinsic EF sources considered in this work arise from realistic coil misalignments in ITER.
We consider rigid shifts and tilts of the PF, TF, and CS coil systems, with amplitudes chosen to be representative of engineering tolerances~\cite{GribovIAEA12,Pharr2024,Bai2025}.
Each EF source is treated independently by computing the perturbation produced by a unit misalignment of the corresponding coil set.
For the primary quantitative comparisons reported below, the unit amplitude is $x=1$~cm and denotes a 1~cm rigid coil shift. Coil tilts are specified separately as angular perturbations about the coil center in the GPEC geometry input. The deformation in Fig.~\ref{fig:pf4tilt-pf3shift} is deliberately exaggerated and is not drawn to the numerical perturbation amplitude.
Because the response model is formulated linearly in the applied spectra, the external spectrum of a given misalignment scales with $x$, and configurations with multiple simultaneous misalignments can be constructed by linear superposition of the individual source spectra.
Resonant quantities are accordingly reported per unit amplitude (e.g. $|\tilde{\Phi}_r|$ in G\,cm$^{-1}$), whereas NTV torques, which are quadratic in the perturbation, are reported normalized by $x^2$ (N\,m\,cm$^{-2}$). The physical values for a given misalignment follow by multiplying by $x$ and $x^2$, respectively.
The complete calculation set includes PF/CS/TF shifts and tilts. PF shifts are plotted as a compact representative set below. The other source classes and tilt cases exhibit the same qualitative ordering of the correction strategies.
Figure~\ref{fig:pf4tilt-pf3shift} illustrates two misalignment cases, shown with exaggerated deformation for visualization.

\begin{figure}[t]
  \centering
  \includegraphics[width=\linewidth]{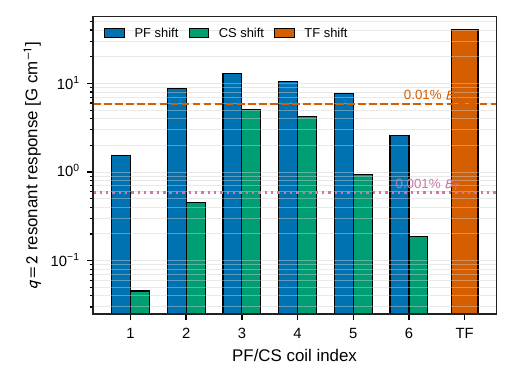}
  \caption{Magnitude of the intrinsic $n=1$, $q=2$ resonant response in the 15~MA baseline equilibrium, normalized per centimeter of rigid coil shift. Dashed and dotted lines show 0.01\% and 0.001\% of $B_T$, respectively.}
  \label{fig:onlyEF-shift-resonance}
\end{figure}

\begin{figure}[t]
  \centering
  \includegraphics[width=\linewidth]{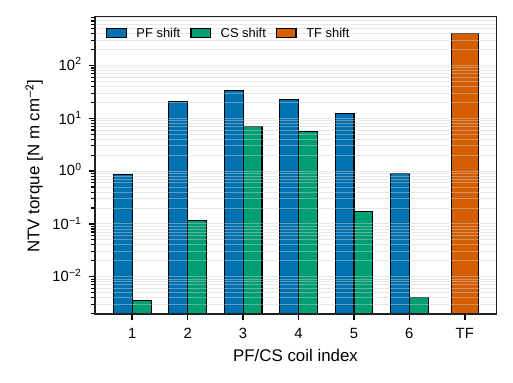}
  \caption{Intrinsic NTV torque in the 15~MA baseline equilibrium for the same rigid-shift sources as Fig.~\ref{fig:onlyEF-shift-resonance}, normalized by the square of the displacement.}
  \label{fig:onlyEF-shift-torque}
\end{figure}

To quantify the impact of these fields on the plasma, we evaluate both the resonant response and the NTV torque.
The resonant response is characterized using the complex resonant flux on the selected $n=1$ rational surfaces.
Figure~\ref{fig:onlyEF-shift-resonance} summarizes the resonant response produced by unit shifts of PF, CS, and TF coils, evaluated on the same rational surfaces.
The nonresonant transport impact is quantified by the NTV torque computed by GPEC, and the corresponding torque response for the same unit shifts is summarized in Fig.~\ref{fig:onlyEF-shift-torque}.

The plasma response to nonaxisymmetric magnetic perturbations is computed using GPEC~\cite{ParkJKPOP07,Park2017}.
For each EF source and each correcting coil, the external spectrum evaluated on the control surface is supplied as the GPEC boundary input.
GPEC returns the shielded response, from which we extract the complex resonant flux on the selected rational surfaces.
We collect these resonant quantities into a resonant-flux vector $\tilde{\Phi}_r$ by concatenating the complex resonant flux evaluated on the $q=2$ and $q=3$ surfaces.
By assembling responses across multiple actuator channels and EF sources, we form a resonant coupling matrix that relates applied spectra to the resonant response.
SVD of this matrix is used to identify dominant response structures, which describe the resonant drive responsible for locking and form the basis for overlap correction.
Both $q=2$ and $q=3$ enter the optimization. We report $q=2$ in the primary comparison because it is the core surface most prone to locking in the present equilibria. The transferred responses of both surfaces are shown in Sec.~\ref{sec:robustness}.

In the primary comparisons, we restrict the actuator set to the three EFC coil families EFCB, EFCM, and EFCT.
Although the actuator space can be expanded by including RMP coils, which can improve correction efficiency, we reserve the RMP capability for the control of edge-localized modes (ELMs)~\cite{LoarteNF14,ParkJKRMPP23} and focus on EFC-only correction when comparing strategies.
Within this EFC-only space, we treat overlap correction as the operational reference, resonance-optimal correction as a conditional extension when sufficient actuator freedom is available, and QS EFC as the method that uses the remaining freedom to minimize torque after resonant acceptability has been secured.

\begin{figure}[t]
  \centering
  \includegraphics[width=\linewidth]{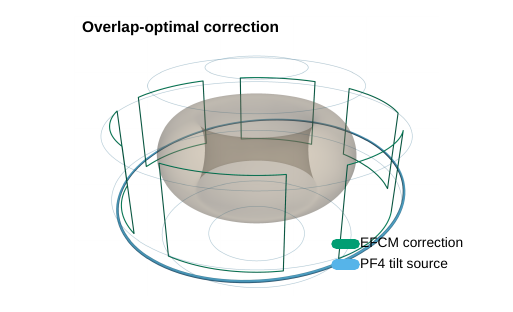}
  \caption{Schematic illustration of overlap-optimal correction. A representative PF4-tilt source is countered primarily with EFCM to reduce its projection onto the dominant resonant-response structure. The geometric displacement is exaggerated for visibility.}
  \label{fig:overlap-cor-3d}
\end{figure}

The first strategy is overlap-optimal correction~\cite{Park2008,Pharr2024}, which targets the dominant resonant structure identified by the SVD of the resonant coupling matrix.
In this approach, the correcting currents are chosen to minimize the projection of the residual resonant response onto the dominant left singular vector.
This corresponds closely to the conventional dominant-mode nulling concept and is achievable with a single coil family in ITER, which is why it is commonly associated with EFCM-only correction (Fig.~\ref{fig:overlap-cor-3d}).

\begin{figure}[t]
  \centering
  \includegraphics[width=\linewidth]{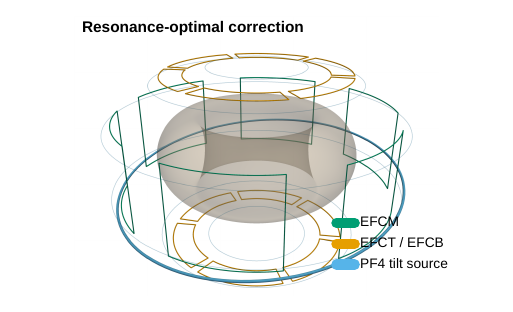}
  \caption{Schematic illustration of resonance-optimal correction for the same PF4-tilt source shown in Fig.~\ref{fig:overlap-cor-3d}. EFCB, EFCM, and EFCT are combined to minimize the residual resonant-flux vector on the selected rational surfaces.}
  \label{fig:nrmp-cor-3d}
\end{figure}

The second strategy is resonance-optimal correction, which conditionally extends overlap-optimal correction. This method minimizes the full residual resonant flux on the selected rational surfaces.
The approach is natural in the present equilibria because the dominant $n=1$ rational surfaces are limited to $q=2$ and $q=3$, so the resonant target is of low dimension.
Using three EFC coil families (Fig.~\ref{fig:nrmp-cor-3d}) gives enough degrees of freedom to reduce the full two-component resonant-flux vector. The optimization then suppresses the dominant SVD mode together with the sub-dominant components that contribute to the total residual resonance.

The third strategy is QS EFC based on the QSMP concept introduced in Sec.~\ref{sec:torque-matrix}.
In this approach, the correcting currents are chosen to minimize the NTV torque computed by GPEC, subject to a resonance cap that keeps the residual resonance below a threshold set by field penetration~\cite{LoganPPCF20}.
The aim is to keep resonance control reliable and to spend the remaining degrees of freedom on reducing the quasi-symmetry breaking that drives nonresonant braking.

All three strategies can be expressed within a unified linear-response model for the resonant flux.
With the correcting-coil currents $\mathbf{I}_c$, the EF-source amplitudes $\mathbf{I}_s$, and the augmented vector $\mathbf{I}$ of Eq.~(\ref{eq:torque-block}), we model the resonant-flux vector on the selected rational surfaces by linear superposition,
\begin{equation}
\begin{split}
\tilde{\Phi}_r(\mathbf{I}_c;\mathbf{I}_s)
&\approx \mathbf{C}_c\,\mathbf{I}_c
 + \mathbf{C}_{EF}\,\mathbf{I}_s
 = \mathbf{C}_{\mathrm{tot}}\,\mathbf{I},\\
\mathbf{C}_{\mathrm{tot}}
&\equiv [\,\mathbf{C}_c\;\;\mathbf{C}_{EF}\,].
\end{split}
\label{eq:Phi_linear}
\end{equation}
Here $\mathbf{C}_c$ maps correcting-coil currents to the resonant response and $\mathbf{C}_{EF}$ maps EF sources to the same response vector.
In practice, the EF realization $\mathbf{I}_s$ is fixed by the misalignment scenario and only $\mathbf{I}_c$ is optimized.
Actuator limits are enforced coil by coil through $|I_{c,k}| \le I_{\max,k}$, where $I_{\max,k}$ is the engineering limit of coil family $k$ (Sec.~\ref{sec:economy}).

Overlap-optimal correction is formulated by selecting the dominant left singular vector $\mathbf{u}_1$ of the resonant coupling operator and minimizing the dominant-mode projection of the residual resonant response,
\begin{equation}
\min_{\mathbf{I}_c}\;\;
\left| \mathbf{u}_1^\dagger \tilde{\Phi}_r(\mathbf{I}_c;\mathbf{I}_s) \right|^2
\quad \text{s.t.}\quad
|I_{c,k}| \le I_{\max,k}\;\;\forall k.
\label{eq:opt_overlap}
\end{equation}

Resonance-optimal correction is formulated by minimizing the total residual resonance across the selected rational surfaces,
\begin{equation}
\min_{\mathbf{I}_c}\;\;
\left\| \tilde{\Phi}_r(\mathbf{I}_c;\mathbf{I}_s) \right\|_2^2
\quad \text{s.t.}\quad
|I_{c,k}| \le I_{\max,k}\;\;\forall k.
\label{eq:opt_resonance}
\end{equation}

For the simplified target used here, the three complex EFC currents are constrained by two complex resonant targets, so one complex degree of freedom generally remains after exact target cancellation. The resonance-optimal currents are hence not unique. The results reported below use the member with the smallest peak coil current relative to the engineering limits, found by moving along the null space of $\mathbf{C}_c$ from the minimum-norm solution. No torque criterion is applied in this selection, so the member should not be read as either the best or the worst torque solution within the resonantly equivalent family.

Resonance-optimal correction extends overlap nulling only in a certain regime. The resonant constraints are counted by the selected rational surfaces and the poloidal harmonics retained in $\tilde{\Phi}_r$. If they outnumber the actuator degrees of freedom, multi-surface cancellation is overdetermined. Overlap correction is then still the sound and economical choice, since it spends the limited actuation on the most amplified response structure identified by the SVD.
The present case is the opposite. Three independent EFC coil families act on an $n=1$ target dominated by the $q=2$ and $q=3$ surfaces, so the actuator space exceeds the target dimension. The residual resonance after dominant-mode nulling is then set by sub-dominant singular structures, and minimizing $\|\tilde{\Phi}_r\|_2$ suppresses the dominant and sub-dominant components together.
A limiting example is \emph{EFCM-only} correction, where the actuator is effectively one-dimensional. The overlap metric is then the natural prescription because it selects the single current phase that best cancels the dominant mode. Full resonance minimization is not feasible without additional independent coil families~\cite{Park2012}.

QS EFC based on QSMP is formulated by minimizing the NTV torque while constraining the residual resonance below a cap,
\begin{equation}
\begin{split}
\min_{\mathbf{I}_c}\quad
&T(\mathbf{I}_c,\mathbf{I}_s)\\
\text{s.t.}\quad
&\left\| \tilde{\Phi}_r(\mathbf{I}_c;\mathbf{I}_s) \right\|_2
 \le \Phi_{\mathrm{cap}},\\
&|I_{c,k}| \le I_{\max,k}\;\;\forall k.
\end{split}
\label{eq:opt_qsmp}
\end{equation}
The cap used for the torque-aware optimizations is $\Phi_{\mathrm{cap}}=0.001\%\,B_T=10^{-5}B_T$.
The torque is minimized numerically from multiple initial currents with the source amplitude fixed to unity. In every case reported here, the solution obtained with the current limits alone already had a residual resonance below this level, so the cap remained inactive.
Accordingly, the reported QS EFC solutions are determined by torque minimization rather than by an active resonance constraint.

The cap is retained as a safeguard for other EF realizations or off-nominal equilibria, where a torque-minimizing solution could in principle degrade resonant control.
It imposes the threshold requirement on penetration and locking without affecting the solutions reported here. Torque and rotation damping are treated as continuous performance penalties.

Because disruptive EF scans are not acceptable at $I_p=15$~MA, the optimization is performed at $I_p=7.5$~MA and the resulting correction currents are applied to the 15~MA baseline equilibrium without re-optimization.
By comparing resonance and torque at both current levels, we assess whether low-current identification followed by full-current correction is feasible and how this depends on the correction strategy.

The rationale for this 7.5~MA to 15~MA transfer is that the EF provenance and the dominant response structure should not change dramatically under current scaling.
The two equilibria share $q_{95}<4$ and the same dominant $n=1$ rational surfaces, $q=2$ and $q=3$ (Fig.~\ref{fig:iter-bsl-eqinfo}). The EF sources are fixed by coil geometry and not by the plasma, so the external spectra on the control surface and the dominant singular structure of the resonant coupling matrix are expected to remain similar between the two current levels.
Under these conditions, correction currents optimized to suppress resonance at 7.5~MA remain aligned with similar response directions at 15~MA, which leads to retention of low residual resonance.
Whether the torque advantage of QS EFC is likewise preserved under the transfer is examined in Sec.~\ref{sec:robustness}.

\section{Resonance and torque: comparison of correction strategies}
\label{sec:results}

In this section, we compare the operational overlap reference, the conditional multimodal resonance-optimal extension, and QS EFC using two complementary outcome measures.
The first is the residual resonant response on the selected $n=1$ rational surfaces, which is tied to penetration and locking risk in resonance-centric EFC practice.
The second is the residual NTV torque, which represents nonresonant quasi-symmetry breaking and the associated rotation damping.
Our goal is to determine whether QS EFC reduces the residual torque more than the resonance-focused strategies while keeping the resonant response acceptably low.
The calculation set includes PF, CS, and TF shifts and tilts. The same qualitative ordering is found across the source classes, and PF shifts are presented here as a compact representative set.

\begin{figure*}[t]
  \centering
  \includegraphics[width=\textwidth]{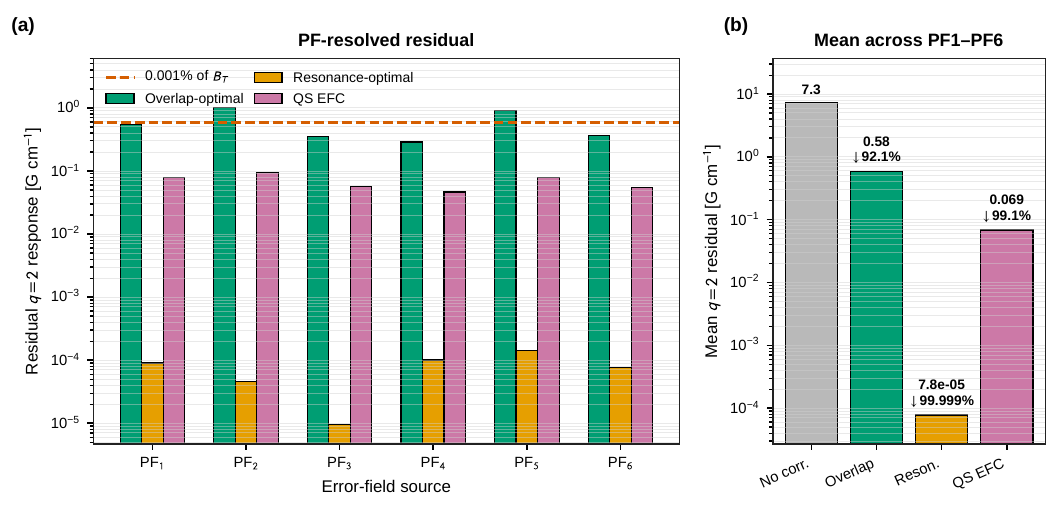}
  \caption{Resonant-field correction performance for 1~cm PF shifts in the 15~MA baseline equilibrium. Panel (a) shows the residual $n=1$, $q=2$ response in G\,cm$^{-1}$ for each PF source. The dashed line is the $0.001\%\,B_T$ cap. Panel (b) compares the PF1--PF6 mean with the intrinsic, uncorrected mean. The QS EFC bars are obtained from torque minimization. The resonance cap is inactive for these solutions.}
  \label{fig:compare-residual-resonance}
\end{figure*}

Figure~\ref{fig:compare-residual-resonance} compares the residual resonant response after applying each correction strategy.
All three strategies reduce the resonant metric far below the intrinsic level, so each can suppress the resonant drive for the tested misalignment set.
As expected from its definition, the resonance-optimal strategy yields the smallest numerical residual because it minimizes the resonant-flux norm itself on the selected rational surfaces.
The overlap-optimal strategy also reduces resonance strongly, but it targets only the dominant resonant structure.
As a result, sub-dominant singular components can remain partially excited. QS EFC produces a smaller $q=2$ residual than overlap correction for every PF source shown.

If an equilibrium contains many relevant rational surfaces while actuator degrees of freedom remain limited, exact multi-surface resonance cancellation becomes difficult and the dominant-mode overlap metric can serve as a dependable, low-parameter correction principle.
In the present ITER equilibria, however, the dominant $n=1$ resonant targets are essentially limited to the $q=2$ and $q=3$ surfaces, so several strategies can drive the residual resonance far beneath the comparison levels used here.
Strategies must therefore be distinguished by additional physics metrics such as NTV torque.

QS EFC is formulated differently, since it minimizes torque under a resonance cap and does not minimize resonance itself. Its mean $q=2$ residual of $6.9\times10^{-2}$~G\,cm$^{-1}$ lies three orders of magnitude above the resonance-optimal value of $7.8\times10^{-5}$~G\,cm$^{-1}$, as expected when a two-component target is cancelled exactly by three actuators. The QS EFC mean nevertheless remains well below the $0.001\%\,B_T$ cap and below 1\% of the intrinsic mean. Because field penetration is threshold-like, reducing an already sub-threshold residual by further orders of magnitude is operationally secondary, whereas NTV torque remains a continuous performance penalty. With three EFC coil families available, QS EFC uses the remaining freedom to form a low-torque perturbation once the resonant requirement has been satisfied.

\begin{figure*}[t]
  \centering
  \includegraphics[width=\textwidth]{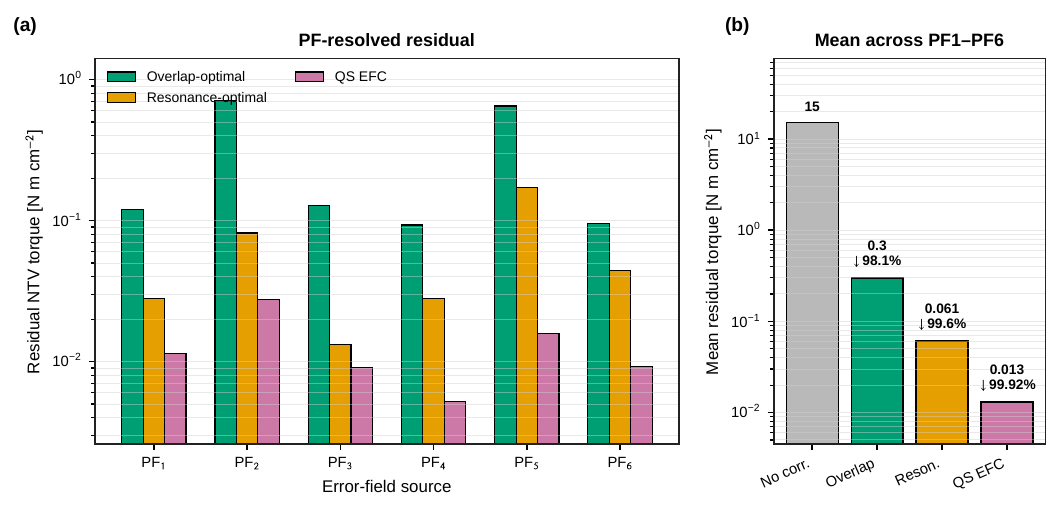}
  \caption{NTV-torque correction performance in the 15~MA baseline equilibrium for the same 1~cm PF shifts and strategies as Fig.~\ref{fig:compare-residual-resonance}. Panel (a) gives the PF-resolved residual torque and panel (b) compares PF1--PF6 means with the intrinsic, uncorrected mean. QS EFC gives the lowest residual for every tested PF source.}
  \label{fig:compare-residual-torque-corrected}
\end{figure*}

Figure~\ref{fig:compare-residual-torque-corrected} compares the residual NTV torque obtained under each strategy.
Suppressing resonance does not, by itself, control the torque.
Resonance-focused corrections do reduce the torque relative to the intrinsic EF in most cases, because both metrics originate from the same nonaxisymmetric spectrum, but the reduction is a by-product rather than an optimization target. It varies widely from source to source, and for several PF sources the residual torque of the overlap-optimal or resonance-optimal solution exceeds that of the QS EFC solution by an order of magnitude or more.
Nothing in the resonant objective prevents the complementary, nonresonant part of the spectrum from being enhanced. In principle, a null-space component can change, and potentially increase, the torque while leaving the resonant objective unchanged. The resonance-optimal solution used here was not selected by a torque criterion. Section~\ref{sec:robustness} shows that its transferred PF6 torque rises above the intrinsic level.
Resonance-focused optimization thus redistributes the three-dimensional spectrum to cancel resonant components on the targeted surfaces, without regard to the nonresonant symmetry-breaking content that drives NTV braking.

\begin{figure}[t]
  \centering
  \includegraphics[width=\linewidth]{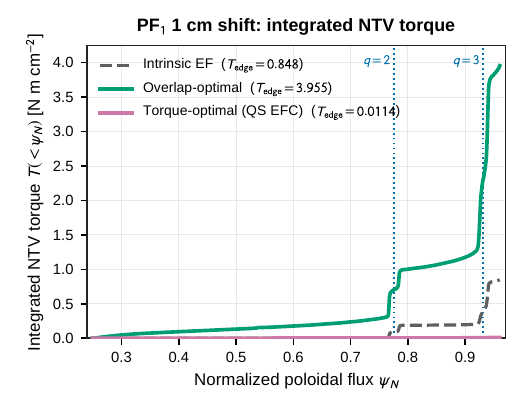}
  \caption{Radial accumulation of the GPEC integrated NTV torque, $T(<\psi_N)$, for a 1~cm PF1 shift in the 15~MA baseline equilibrium. The intrinsic EF is shown in gray, the full three-family dominant-overlap optimum in green, and QS EFC in purple, consistent with the strategy colors used throughout. The dotted vertical lines mark the $q=2$ and $q=3$ surfaces. Although the overlap solution nearly nulls the first SVD component, amplification of the second component raises its edge-integrated torque above the intrinsic value. This diagnostic full-space overlap optimum is distinct from the EFCM-only operational overlap reference used in the aggregate strategy comparisons.}
  \label{fig:pf1-overlap-null-torque}
\end{figure}

Figure~\ref{fig:pf1-overlap-null-torque} gives a direct example. Nulling the single dominant overlap component with all three EFC families leaves two complex degrees of freedom, so overlap-optimal solutions form a large family, of which the EFCM-only solution used in Figs.~\ref{fig:compare-residual-resonance} and~\ref{fig:compare-residual-torque-corrected} is the operational member. For the PF1 shift, a three-family dominant-overlap solution reduces the first SVD-overlap component to $1.13\times10^{-4}$ of its intrinsic value while increasing the second component to 4.44 times its intrinsic value.
The solution stays within the engineering current limits, with a peak current utilization of 0.98, but its total integrated torque $T(<\psi_N{=}1)$ increases from 0.848 to 3.955~N\,m\,cm$^{-2}$, a factor of 4.66, whereas QS EFC reduces it to 0.0114~N\,m\,cm$^{-2}$. Nulling the dominant overlap alone constrains neither the sub-dominant resonant structure nor the nonresonant spectrum that drives NTV.

QS EFC avoids this outcome because the torque itself is the objective, and minimizing the quadratic torque functional steers the solution toward perturbations that lie closer to quasi-symmetric directions in the actuator space.
As noted in Sec.~\ref{sec:methods}, the resonance cap was never active for the cases reported here, so the low residual resonance of the QS EFC solutions is a property of the torque minimum itself rather than a consequence of the constraint.
As a result, QS EFC produces low-residual-torque solutions across the whole tested EF source set, while its $q=2$ residual is lower than that of overlap correction for all PF shifts shown and lies comfortably below the prescribed cap.

These results point to a trade-off between resonance and torque in the resonance-focused strategies.
When the resonant objective is minimized aggressively, the solution can move toward current directions that reduce the resonant metric efficiently but leave the nonresonant torque far from its minimum.
Better resonant suppression can then coincide with higher residual torque, which is undesirable when rotation damping limits performance.
QS EFC treats resonance as a feasibility constraint and minimizes torque within the feasible region.
Its solutions stay within the resonant bound and give the lowest torque for every tested PF source.

The torque-coupling matrix (Fig.~\ref{fig:torque-matrix}) helps explain these trends.
For each misalignment source, the matrix identifies which coil families offer strong leverage for torque reduction and which polarity or phasing is favorable.
QS EFC solutions tend to align correcting currents with these torque-reducing directions.
Resonance-focused strategies can instead select current directions that cancel the resonant metric on the targeted surfaces but couple weakly to torque reduction. They can even raise the torque when the remaining spectrum amplifies quasi-symmetry breaking.
This explains the cases in which resonance is strongly suppressed but the residual torque stays comparatively large.

The absolute size of a PF-only residual should not be read as the complete ITER torque budget. For example, the intrinsic TF-shift result in Fig.~\ref{fig:onlyEF-shift-torque} is approximately $4\times10^{2}$~N\,m\,cm$^{-2}$, corresponding to about 4~N\,m for a 1~mm shift under quadratic displacement scaling. An actual EF is a coherent superposition of PF, CS, and TF shifts and tilts, and the full torque matrix retains their cross terms. The small direct-correction values in Fig.~\ref{fig:compare-residual-torque-corrected} show the effectiveness of QS EFC for the isolated PF benchmarks, not a global upper bound on residual torque in the assembled machine.

\FloatBarrier
\section{Correction economy and current requirements}
\label{sec:economy}

QS EFC can require larger correcting-coil currents in some cases, depending on the EF source.
In this section, we quantify the required current magnitudes and distributions to assess correction economy and actuator headroom.
The engineering limits used throughout are 320~kA-turn for EFCB and EFCT and 200~kA-turn for EFCM, corresponding to the 10~kA conductor current with 32 and 20 turns, respectively~\cite{WeiIAEA10,AmoskovPPNL15}.

\begin{figure}[t]
  \centering
  \includegraphics[width=\linewidth]{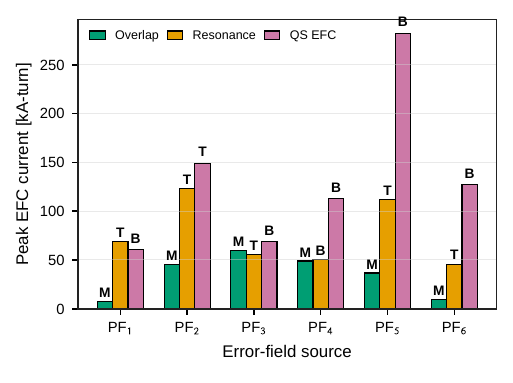}
  \caption{Peak EFC-current demand in kA-turn in the 15~MA baseline equilibrium for 1~cm PF-shift sources. The letter above each bar identifies the dominant row: B, M, and T denote EFCB, EFCM, and EFCT. QS EFC can require more current because torque, rather than current economy, is its objective.}
  \label{fig:current-compare1}
\end{figure}

\begin{figure}[t]
  \centering
  \includegraphics[width=\linewidth]{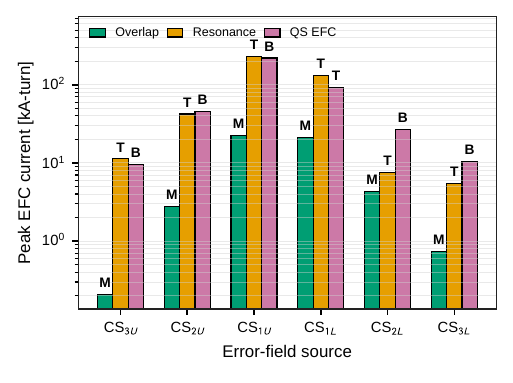}
  \caption{Peak EFC-current demand in kA-turn in the 15~MA baseline equilibrium for 1~cm CS-shift sources. B, M, and T denote the dominant EFCB, EFCM, and EFCT rows. The logarithmic ordinate exposes the broad source-to-source variation in required current.}
  \label{fig:current-compare2}
\end{figure}

\begin{figure*}[t]
  \centering
  \includegraphics[width=\textwidth]{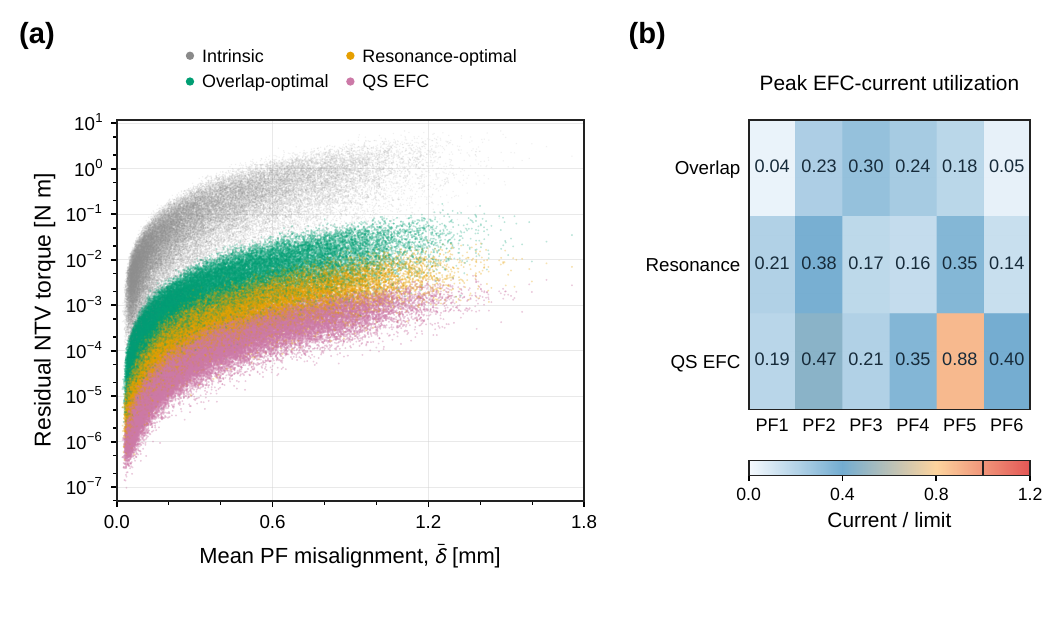}
  \caption{Correction performance and current demand in the 15~MA baseline equilibrium. (a) Residual integrated NTV torque in N\,m for 50\,000 random samples of simultaneous horizontal shifts of PF1--PF6, shown on a linear abscissa and a logarithmic ordinate. The abscissa is the realized mean displacement magnitude across the six PF coils in mm. The expected displacement scale is sampled log-uniformly from 0.05 to 1~mm, and the horizontal displacement components are independent zero-mean Gaussian variables conditional on that scale. The same samples are used for the intrinsic case and the three strategies, and the correction currents are recalculated for each sample. QS EFC uses the minimum of Eq.~(\ref{eq:Ic-opt}) with the source held fixed. The source distribution is illustrative. (b) Peak EFC-current demand divided by the corresponding engineering limit for the separate single-PF 1~cm (10~mm) shift benchmarks. The black mark on the color scale indicates unity. All cases remain below unity, with the largest utilization of approximately 0.88 occurring for QS EFC of the PF5 source. Panel (b) does not represent the ensemble in panel (a).}
  \label{fig:pf-correction-tradeoff}
\end{figure*}

Figures~\ref{fig:current-compare1} and~\ref{fig:current-compare2} show that QS EFC and resonance-optimal correction generally require a larger peak current than overlap-optimal solutions. The letter above each bar (B/M/T, denoting EFCB, EFCM, and EFCT) identifies the coil family that requires the largest correcting current.
This increase has a clear structure. The strongly coupled coil family, typically EFCM for PF-shift sources, provides the bulk correction, while the weakly coupled families supply the remaining fine-tuning. Because their leverage is low, these fine-tuning families need a large current to produce the required effect, so the peak current usually appears on the weakly coupled rows rather than on the strongly coupled one.
QS EFC can thus be more expensive in peak current, which is the cost of spending the extra degrees of freedom on torque reduction rather than on current economy. Because the response is linear in source amplitude, the required correction currents scale linearly with displacement, and the demands plotted for the deliberately conservative 1~cm benchmark shrink in direct proportion for millimeter-scale tolerances. Feedforward estimates for the expected ITER tolerances indeed give required EFC currents of a few tens of kA-turn~\cite{Bai2025}.

Figure~\ref{fig:pf-correction-tradeoff}(a) extends the single-source comparisons to simultaneous horizontal shifts of PF1--PF6.
For each of 50\,000 random source samples, the correction currents are recalculated from the GPEC response and torque matrices, which retain all source--source and source--corrector cross terms.
The abscissa is the realized mean displacement magnitude across the six PF coils. The expected displacement scale is sampled log-uniformly from 0.05 to 1~mm, with independent zero-mean Gaussian displacement components conditional on that scale. This sampling model is illustrative and is not an estimate of the distribution of manufacturing errors in ITER.
QS EFC uses the minimum of Eq.~(\ref{eq:Ic-opt}) with the source held fixed, and the current limits and the resonance cap are inactive for these solutions. Because Eq.~(\ref{eq:Ic-opt}) is the torque minimum, QS EFC gives the lowest torque for every sample, and the figure shows its margin to the other strategies over the sampled range.

Figure~\ref{fig:pf-correction-tradeoff}(b) retains the comparison of current utilization for the single-PF 1~cm shift benchmarks.
All cases remain below the assumed engineering limits. Among the three strategies, the maximum utilization is approximately 0.88, occurring for the PF5 QS EFC case.
This panel describes the single-source benchmarks and not the ensemble in panel (a).

\FloatBarrier
\section{Robustness of low-current optimization}
\label{sec:robustness}

We evaluate the robustness of applying correction currents optimized at reduced current ($I_p=7.5$~MA) directly to the 15~MA baseline without re-optimization.
For this test we compare resonance-optimal correction and QS EFC only, because a resonance-optimal solution that cancels the full resonant-flux vector also nulls its projection onto the dominant singular structure. It therefore satisfies the overlap-optimal condition as well and represents the resonance-focused family in its strongest form.
As in Sec.~\ref{sec:results}, PF shifts are shown.

\begin{figure*}[t]
  \centering
  \includegraphics[width=\textwidth]{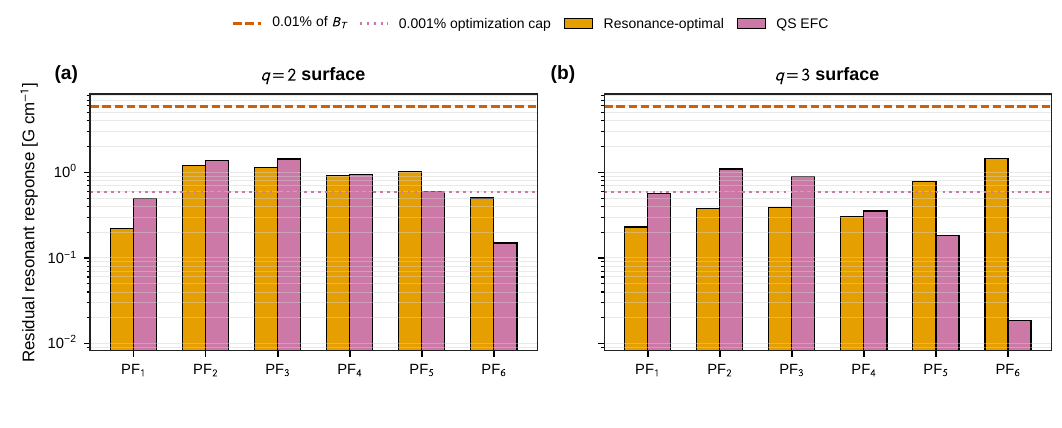}
  \caption{Residual resonant response at $I_p=15$~MA when correction currents optimized at $I_p=7.5$~MA are applied to the 15~MA baseline without re-optimization, for 1~cm PF shifts. Panels (a) and (b) show the $q=2$ and $q=3$ surfaces. The dotted 0.001\%-$B_T$ line is the torque-aware optimization cap at 7.5~MA. It is shown here as a reference after transfer and is not re-imposed at 15~MA. The dashed 0.01\%-$B_T$ line marks the full-current comparison level.}
  \label{fig:7.5toIBSresonance}
\end{figure*}
\begin{figure}[t]
  \centering
  \includegraphics[width=\linewidth]{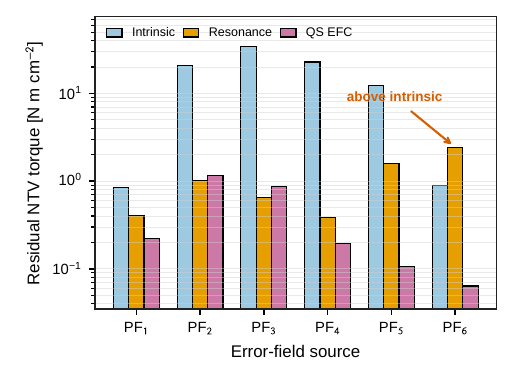}
  \caption{Residual NTV torque at $I_p=15$~MA for the same transfer as in Fig.~\ref{fig:7.5toIBSresonance}. Light-blue bars show the intrinsic 15~MA torque. The annotation marks PF6, for which transferred resonance-optimal correction leaves a torque above the intrinsic level, whereas QS EFC remains below it.}
  \label{fig:7.5toIBStorque}
\end{figure}

At $I_p=7.5$~MA, both strategies achieve meaningful resonance suppression, while QS EFC gives the more consistent torque reduction (not shown). After the 7.5~MA-optimized currents are transferred to the 15~MA equilibrium (Figs.~\ref{fig:7.5toIBSresonance} and~\ref{fig:7.5toIBStorque}), the $q=2$ and $q=3$ residuals of both strategies rise to values of order 1~G\,cm$^{-1}$ and are not negligible relative to their low-current optima. Nevertheless, even the largest transferred value is still several times below the $0.01\%\,B_T$ comparison level.

The transferred torques are likewise of order 1~N\,m\,cm$^{-2}$ for several sources, and QS EFC is not smaller source by source, as its residual exceeds the resonance-optimal value for PF2 and PF3. The mean, however, is 0.44~N\,m\,cm$^{-2}$ for QS EFC and 1.08~N\,m\,cm$^{-2}$ for resonance-optimal correction, a factor of about 2.5. In addition, every transferred QS EFC case remains below its intrinsic torque, whereas the resonance-optimal PF6 case rises above the intrinsic value.

The different local sensitivities of the two optima help interpret this result. For a fixed target equilibrium, expansion of the torque about a QS EFC optimum $\mathbf{I}_c^\ast$ has no first-order current-mismatch term because $\nabla_{\mathbf{I}_c}T(\mathbf{I}_c^\ast)=0$. The leading current-only penalty is therefore $O(\lVert\delta\mathbf{I}_c\rVert^2)$. A representative resonance-optimal solution is generally not stationary with respect to torque, so its torque variation can contain an $O(\lVert\delta\mathbf{I}_c\rVert)$ contribution. The equilibrium and response operator also change during the 7.5-to-15~MA transfer, so QS EFC still degrades in both metrics. The stationarity argument explains the reduced local torque sensitivity but does not imply invariance under transfer.

These results support an ITER pathway of \emph{low-current identification} combined with QS EFC.
Such a workflow avoids disruptive full-current scans, retains a resonance margin, and reduces the mean NTV torque after transfer.

\FloatBarrier
\section{Conclusions}
\label{sec:conclusion}

This paper developed and assessed QS EFC for intrinsic EF correction in ITER, with workflows compatible with the 15~MA baseline scenario. GPEC plasma-response calculations were carried out for PF/TF/CS coil shifts and tilts in ITER, and three strategies were compared. The first is overlap-optimal dominant-mode correction, the operational reference. The second is a conditional resonance-optimal extension that minimizes the total residual resonant flux on selected $n=1$ rational surfaces. The third is QS EFC based on the QSMP concept, which minimizes the NTV torque while retaining a resonance safeguard against field penetration.

The broader implication for reactor-scale operation is that torque effects must be treated on an equal footing with resonance. In large devices such as ITER, and even more so in next-step machines such as the Burning plasma Experimental Superconducting Tokamak (BEST) and DEMO, rotation control margins are limited, and the performance impact of NTV braking can be disproportionately severe. This follows from the different metric scalings. The resonant response scales approximately \emph{linearly} with the applied nonaxisymmetric field, whereas the NTV torque scales \emph{quadratically}, so modest spectral mismatches that appear tolerable in resonance can translate into large penalties in torque and rotation damping.

Three conclusions follow.

First, the \emph{torque-coupling matrix} makes torque minimization applicable in practice. It maps each EF source to the actuator choice and phasing that reduce its torque. It shows which coil families are most efficient for a given source and why certain actuators repeatedly emerge, such as EFCM for PF-shift sources.

Second, resonance-focused corrections do not uniquely control the residual NTV torque. In the present three-actuator, two-target problem, resonance cancellation leaves a complex degree of freedom, and the representative resonance-optimal solution is not selected by torque. QS EFC gives the lowest torque for every PF-shift case optimized at full current and a smaller $q=2$ residual than overlap correction for every PF source shown. Its mean $q=2$ residual is not numerically as small as the resonance-optimal value, but it lies well below the conservative $0.001\%\,B_T$ cap and below 1\% of the intrinsic mean. Thus, once the threshold-like resonant requirement is met, QS EFC uses the remaining freedom to reduce the continuous torque penalty.

Third, correction currents optimized at reduced plasma current ($I_p=7.5$~MA) retain useful performance when transferred directly to the 15~MA baseline equilibrium without re-optimization. Both strategies degrade to order-1~G\,cm$^{-1}$ residuals, but retain a clear margin to the $0.01\%\,B_T$ comparison level. QS EFC does not give the lower torque for every individual transferred source, but its mean torque is about 2.5 times lower than that of the resonance-optimal solution and all QS EFC cases remain below their intrinsic torques. The vanishing first-order torque gradient at a QS EFC optimum provides a local mechanism for this reduced sensitivity, while not eliminating changes caused by the transferred equilibrium itself.

Sensitivity to profiles, collisionality, rotation, and equilibrium variation should next be quantified to bound uncertainty in the predicted torque and resonance. Extending the framework to multiple toroidal mode numbers~\cite{Paz-SoldanNF15,LanctotPOP17,LoganNF20} and experimentally constrained EF inference will further strengthen QS EFC planning for ITER and reactor-relevant devices.

\section*{Data availability statement}
The data and optimization scripts supporting the findings are available from the corresponding author upon reasonable request.

\section*{Acknowledgments}
We thank collaborators and colleagues for discussions on QSMP, torque-aware optimization, and ITER EFC workflows. This work was supported by the National Research Foundation of Korea (NRF), Ministry of Science and ICT, Nos. RS-2024-00409564 and RS-2024-00350293.

\section*{References}
\bibliographystyle{iopart-num}
\bibliography{ref_iop}

\end{document}